\overfullrule=0pt
\input harvmac
\input amssym
\def\a{{\alpha}}
\def\ah{{\widehat\alpha}}

\def\l{{\lambda}}
\def\lh{{\widehat\lambda}}

\def\b{{\beta}}
\def\bh{{\widehat\beta}}
\def\g{{\gamma}}
\def\gh{{\widehat\gamma}}
\def\d{{\delta}}
\def\dh{{\widehat\delta}}
\def\e{{\epsilon}}
\def\s{{\sigma}}
\def\r{{\rho}}
\def\sh{{\widehat\sigma}}
\def\rh{{\widehat\rho}}
\def\N{{\nabla}}
\def\Nb{{\overline\nabla}}

\def\O{{\Omega}}
\def\Ob{{\overline\O}}
\def\Oh{{\widehat\O}}
\def\o{{\omega}}
\def\oh{{\widehat\omega}}

\def\ve{{\varepsilon}}
\def\veh{{\widehat\varepsilon}}

\def\p{{\partial}}
\def\pb{{\overline\partial}}
\def\t{{\theta}}
\def\th{{\widehat\theta}}
\def\ph{{\widehat p}}
\def\oh{{\widehat\o}}
\def\L{{\Lambda}}

\def\dhh{{\widehat d}}

\def\Pib{{\overline\Pi}}
\def\Jb{{\overline J}}

\def\BB{{\cal B}}

\def\HH{{\cal H}}

\def\dhh{{\widehat d}}
\def\Ch{\widehat C}

\def\Nhh{{\widehat N}}
\def\KK{{\cal K}}
\def\HH{{\cal H}}

\baselineskip12pt

\Title{ \vbox{\baselineskip12pt
}}
{\vbox{\centerline
{ Ambitwistor pure spinor string  }
\bigskip
\centerline{in a type II supergravity background }
}}
\smallskip
\centerline{Osvaldo Chandia\foot{e-mail: ochandiaq@gmail.com}, }
\smallskip
\centerline{\it Departamento de Ciencias, Facultad de Artes Liberales, Universidad Adolfo Ib\'a\~nez}
\centerline{\it Facultad de Ingenier\'{\i}a y Ciencias, Universidad Adolfo Ib\'a\~nez}
\centerline{\it Diagonal Las Torres 2640, Pe\~nalol\'en, Santiago, Chile} 
\bigskip
\centerline{Brenno Carlini Vallilo\foot{e-mail: vallilo@gmail.com}, }
\smallskip
\centerline{\it Departamento de Ciencias F\'{\i}sicas, Facultad de Ciencias Exactas}
 \centerline{\it Universidad Andres Bello, Rep\'ublica 220, Santiago, Chile}

\bigskip
\bigskip
\noindent
We construct the ambitwistor pure spinor string in a general type II supergravity background in the semi-classical regime. Almost all supergravity constraints are obtained from nilpotency of the BRST charge  and further consistency conditions from additional world-sheet 
symmetries. We also discuss the case of $AdS_5\times S^5$ background.

\Date{May 2015}


\lref\MasonSVA{
  L.~Mason and D.~Skinner,
  ``Ambitwistor strings and the scattering equations,''
JHEP {\bf 1407}, 048 (2014).
[arXiv:1311.2564 [hep-th]].
}

\lref\BerkovitsXBA{
  N.~Berkovits,
  ``Infinite Tension Limit of the Pure Spinor Superstring,''
JHEP {\bf 1403}, 017 (2014).
[arXiv:1311.4156 [hep-th], arXiv:1311.4156].
}

\lref\AdamoWEA{
  T.~Adamo, E.~Casali and D.~Skinner,
  ``A Worldsheet Theory for Supergravity,''
JHEP {\bf 1502}, 116 (2015).
[arXiv:1409.5656 [hep-th]].
}

\lref\BerkovitsUE{
  N.~Berkovits and P.~S.~Howe,
  ``Ten-dimensional supergravity constraints from the pure spinor formalism for the superstring,''
Nucl.\ Phys.\ B {\bf 635}, 75 (2002).
[hep-th/0112160].
}

\lref\ChandiaIX{
  O.~Chandia,
  ``A Note on the classical BRST symmetry of the pure spinor string in a curved background,''
JHEP {\bf 0607}, 019 (2006).
[hep-th/0604115].
}

\lref\ChandiaSTA{
  O.~Chandia and B.~C.~Vallilo,
  ``Non-minimal fields of the pure spinor string in general curved backgrounds,''
JHEP {\bf 1502}, 092 (2015).
[arXiv:1412.1030 [hep-th]].
}

\lref\BedoyaIC{
  O.~A.~Bedoya and O.~Chandia,
  ``One-loop Conformal Invariance of the Type II Pure Spinor Superstring in a Curved Background,''
JHEP {\bf 0701}, 042 (2007).
[hep-th/0609161].
}

\lref\CachazoGNA{
  F.~Cachazo, S.~He and E.~Y.~Yuan,
  ``Scattering equations and Kawai-Lewellen-Tye orthogonality,''
Phys.\ Rev.\ D {\bf 90}, no. 6, 065001 (2014).
[arXiv:1306.6575 [hep-th]].
}
\lref\CachazoHCA{
  F.~Cachazo, S.~He and E.~Y.~Yuan,
  ``Scattering of Massless Particles in Arbitrary Dimensions,''
Phys.\ Rev.\ Lett.\  {\bf 113}, no. 17, 171601 (2014).
[arXiv:1307.2199 [hep-th]].
}
\lref\CachazoIEA{
  F.~Cachazo, S.~He and E.~Y.~Yuan,
  ``Scattering of Massless Particles: Scalars, Gluons and Gravitons,''
JHEP {\bf 1407}, 033 (2014).
[arXiv:1309.0885 [hep-th]].
}

\lref\MikhailovRX{
  A.~Mikhailov,
  ``Symmetries of massless vertex operators in AdS(5) x S**5,''
J.\ Geom.\ Phys.\  {\bf 62}, 479 (2012).
[arXiv:0903.5022 [hep-th]].
}

\lref\ChandiaKJA{
  O.~Chandia, A.~Mikhailov and B.~C.~Vallilo,
  ``A construction of integrated vertex operator in the pure spinor sigma-model in $AdS_5 \times S^5$,''
JHEP {\bf 1311}, 124 (2013).
[arXiv:1306.0145 [hep-th]].
}
\lref\WittenNN{
  E.~Witten,
  ``Perturbative gauge theory as a string theory in twistor space,''
Commun.\ Math.\ Phys.\  {\bf 252}, 189 (2004).
[hep-th/0312171].
}
\lref\BerkovitsHG{
  N.~Berkovits,
  ``An Alternative string theory in twistor space for N=4 superYang-Mills,''
Phys.\ Rev.\ Lett.\  {\bf 93}, 011601 (2004).
[hep-th/0402045].
}

\lref\ElvangCUA{
  H.~Elvang and Y.~t.~Huang,
[arXiv:1308.1697 [hep-th]].
}

\lref\GrossAR{
  D.~J.~Gross and P.~F.~Mende,
  ``String Theory Beyond the Planck Scale,''
Nucl.\ Phys.\ B {\bf 303}, 407 (1988).
}

\lref\GrossKZA{
  D.~J.~Gross and P.~F.~Mende,
  ``The High-Energy Behavior of String Scattering Amplitudes,''
Phys.\ Lett.\ B {\bf 197}, 129 (1987).
}

\lref\NathanPrivate{N.~Berkovits, private communication.}

\newsec{Introduction}

For more than a decade, twistor string theories describing massless supersymmetric field theories in four dimensions 
have been formulated \refs{\WittenNN,\BerkovitsHG}. These formulations together with related developments lead to a revived interest in twistor methods which greatly improved the knowledge and techniques to compute scattering amplitudes\foot{The number of relevant papers is far to great to cite here. We recommend the review \ElvangCUA.}. The difficulty in generalizing these strings to higher dimensions lies in the fact that momentum twistors do not generalize to higher dimensions. 

More recently, in a series of interesting papers Cachazo, He and Yuan constructed all tree level amplitudes for a large class of massless theories in any dimension \refs{\CachazoGNA,\CachazoHCA,\CachazoIEA}. The structure of their expressions resembled the previous work of Gross and Mende \refs{\GrossAR,\GrossKZA} on high energy scattering of string theory. Later, Mason and Skinner argued that such amplitudes can be computed using an ambitwistor string \MasonSVA. The action for this string is first order and chiral. It also has more world-sheet symmetries than the usual string. The critical dimenstion for this string is $26$ in the bosonic case and $10$ in the RNS case. The striking difference from the usual strings is that it has no massive states. This is one of the features that allow their amplitudes agree to with those of 
Cachazo {\it et al.} without taking the usual $\alpha'\to 0$ limit. 

Because of the complications with space-time supersymmetry in the RNS formalism, it is natural to use the pure spinor formalism in the supersymmetric case. This was done by Berkovits in \BerkovitsXBA, where the world-sheet action in a flat space-time looks like the usual pure spinor string but with all the fields are holomorphic. One could also ask if it is possible to formulate the Mason-Skinner string in a type II supergravity background. This was done in \AdamoWEA\  for the RNS formulation, but only for the bosonic background fields. The purpose of this paper is to construct the pure spinor formulation in a general supergravity background. We found that in addition to the BRST charge, the curved space formulation requires additional symmetries to obtain almost all supergravity constraints. In particular, the masshell condition $P_aP^a=0$ which follows from BRST in \BerkovitsXBA\ needs to be generalized to include background fields that vanish in the flat space limit. In the present work we consider only the semi-classical limit. As it was shown in 
\AdamoWEA, the full quantum theory requires corrections to the classical constraints. We expect the same to be true in our model. However, as in \AdamoWEA, these quantum corrections are not expected to modify the supergravity constraints. We hope to prove this in a future work. We also discuss the application of our results to the case of $AdS_5\times S^5$. The resulting action is much simpler than the usual case. Furthermore, the global $\frak{psu}(2,2|4)$ symmetry is promoted to full Kac-Moody holomorphic invariance. It is possible that one can use such symmetry to completely solve the model. 

This paper is organized as follows. In section 2 we review the ambitwistor string on a flat space \BerkovitsXBA\ and introduce the role of the $B$ field in this system. In section 3 we construct the curved background version and obtain the type II supergravity constraints for the background after requiring BRST invariance. In section 4 we study the special type IIB background with $AdS_5 \times S^5$ geometry. We end with concluding remarks in section 5. 

\newsec{Flat Background}

In a flat background the world-sheet action is
\eqn\sflat{S=\int d^2z ~ P_m \pb X^m + p_\a \pb \t^\a + \ph_\ah \pb \th^\ah + \o_\a \pb \l^\a + \oh_\a \pb \lh^\ah ,}  
where $(X^m,\t^\a,\th^\ah)$ are the flat ten-dimensional $N=2$ superspace coordinates, and $(P_m, p_\a, \ph_\ah)$ are their momentum conjugates. The pure spinors $\l^\a$ and $\lh^\ah$, and their momentum conjugate variables $\o_\a$ and $\oh_\ah$, enter as in the minimal pure spinor formalism except that all these variables are holomorphic. A BRST-like charge is defined as
\eqn\brst{Q=\oint \l^\a d_\a + \l^\ah \dhh_\ah ,}
where 
\eqn\defd{Êd_\a = p_\a - \ha P_m (\g^m\t)_\a,\quad \dhh_\ah = \ph_\ah - \ha P_m (\g^m\th)_\ah .} The world-sheet fields $(P_m, \l^\a, \lh^\ah)$ are BRST invariant and the remaining fields transform as
\eqn\qbrst{Q\t^\a = \l^\a,\quad Q\th^\ah = \lh^\ah,\quad Q\o_\a = d_\a,\quad Q\oh_\ah = \dhh_\ah ,} 
$$ Q X^m = \ha (\l\g^m\t) + \ha (\lh\g^m\th),\quad Q p_\a = -\ha P_m (\g^m\l)_\a,\quad Q \ph_\ah = -\ha P_m (\g^m\lh)_\ah .$$
The action \sflat\ is invariant under \qbrst.

The action \sflat\ is supersymmetric, although it is not manifest. The fields in \sflat\ transform under supersymmetry as
\eqn\susy{Ê\d\t^\a = \ve^\a,\quad \d\th^\ah = \veh^\ah,\quad \d P_m = 0 ,}
$$ \d X^m = - \ha (\ve\g^m\t) - \ha (\veh\g^m\th),\quad \d p_\a = +\ha P_m (\g^m\ve)_\a,\quad \d \ph_\ah = +\ha P_m (\g^m\veh)_\ah .$$ 
It turns out that $d_\a$ and $\dhh_\ah$ are invariant under supersymmetry. Then we could try to write the action \sflat\ in terms of them. The answer to this is
\eqn\ssusy{S=\int d^2z ~ P_m \Pib^m + d_\a \pb \t^\a + \dhh_\ah \pb \th^\ah + \o_\a \pb \l^\a + \oh_\ah \pb \lh^\ah ,}
where
\eqn\defpi{\Pib^m = \pb X^m + \ha (\t\g^m\pb\t) + \ha (\th\g^m\pb\th) ,}
which is invariant under \susy, and therefore the action \ssusy\ is manifestly supersymmetric. 

In the usual  type II superstring, when the action is expressed in terms of the supersymmetric combinations of $(X, p, \ph, \t, \th)$, the WZ term appears which has the form 
\eqn\BB{Ê\int d^2z  ~ \Pi^A \Pib^B B_{BA} ,}
where $\Pi^A = (\Pi^m, \p\t^\a, \p\th^\ah)$ and the non-zero components of $B$ are
\eqn\compB{ÊB_{m\a} = (\g_m \t)_\a,\quad B_{m\ah} = - (\g_m \th)_\ah,\quad B_{\a\bh} = -\ha (\g^mÊ\t)_\a (\g_m \th)_\bh .}
In curved space background, the action will contain a term \BB\ with a supergravity $B$-field. In our case, we do not obtain the WZ term, therefore we are not going to get a term \BB\ in curved background. Then, where is the $B$ field in the system described by the action \sflat\ and the BRST charge \brst? We will prove now that the action \sflat\ has a new symmetry which involves the $B$ field on a flat space. This new symmetry is generated by
\eqn\qflat{ {\cal K} = \oint \l^\a ( \Pi^m B_{m\a} + \p \th^\bh B_{\bh\a} ) +  \lh^\ah  ( \Pi^m B_{m\ah} + \p \t^\b B_{\b\ah} )}
$$= \oint \left( (\l \g^m \t) - (\lh \g^m \th) \right) \Pi_m - \ha (\l \g^m \t) (\th \g_m \p \th) + \ha (\lh \g^m \th) (\t \g_m \p \t) .$$ 
Note that the integrand in \qflat\ is trivially conserved because the world-sheet fields are all holomorphic. Under ${\cal K}$, the world-sheet fields $(X, \t, \th, \l, \lh)$ are invariant. The other fields transform as
\eqn\KP{ \d P_m = -\p(\l\g_m\t) + \p(\lh \g_m \th) ,}
\eqn\Kp{Ê\d d_\a = (\l \g^m)_\a \Pi_m - \ha (\l \g^m)_\a (\th \g_m \p\th) - (\l \g^m \t) (\g_m \p \t)_\a - \ha \p(\lh \g^m \th) (\g_m \t)_\a ,}
\eqn\Kph{Ê\d \dhh_\ah = - (\lh \g^m)_\ah \Pi_m + \ha (\lh \g^m)_\ah (\t \g_m \p\t) + (\lh \g^m \th) (\g_m \p \th)_\ah + \ha \p(\l \g^m \t) (\g_m \th)_\ah ,}
\eqn\Ko{ \d \o_\a = (\g^m \t)_\a \Pi_m - \ha (\g^m \t)_\a (\th \g_m \p \th) ,}
\eqn\Koh{ \d \oh_\ah = - (\g^m \th)_\ah \Pi_m + \ha (\g^m \th)_\ah (\t \g_m \p \t) .}
The action \ssusy\ is invariant under the above transformations. In fact, the terms involving $(\l, X)$ in the variation of the action are
\eqn\dSone{Ê\int d^2 z ~ - \p(\l \g_m \t) \pb X^m + \pb(\l \g^m \t) \p X_m ,}
which vanishes after integrating by parts. Similarly, the terms involving $(\lh , X)$ are
\eqn\dStwo{\int d^2 z ~ Ê\p(\lh \g_m \th) \pb X^m - \pb(\lh \g^m \th) \p X_m ,}
which also vanishes after integrating by parts. The remaining terms in the variation involving $\l$ are
\eqn\dSthree{\int d^2 z ~ -\ha \p(\l \g^m \t) (\t \g_m \pb \t) + \ha \pb(\l \g^m \t) (\t \g_m \p \t) - (\l \g^m \t) (\p\t \g_m \pb \t) ,}
which vanishes after integrating by parts. Finally, the remaining terms in the variation of the action involving $\lh$ are
\eqn\dSfour{\int d^2 z ~ \ha \p(\lh \g^m \th) (\th \g_m \pb \th) - \ha \pb(\lh \g^m \th) (\th \g_m \p \th) + (\lh \g^m \th) (\p\th \g_m \pb \th) ,}
which also vanishes after integrating by parts. Then, we have proved that the action \sflat\ is invariant under the transformations generated by \qflat. The symmetry generated by \qflat\ is also BRST invariant. A short calculation shows
\eqn\QK{ [Q ,{\cal K}] = \ha \oint (\lh \g^m \th) \p(\l \g_m \t) + \p(\lh \g^m \th) (\l \g_m \t) = 0,}
because the integrand is a total derivative. Here we have used the gamma-matrices identities
\eqn\idg{Ê(\g^m)_{(\a\b} (\g_m)_{\g)\d} = 0,\quad (\g^m)_{(\ah\bh} (\g_m)_{\gh)\dh} = 0 ,}
and the pure spinor conditions $(\l \g^m \l) = (\lh \g^m \lh) = 0$. 

The action \sflat\ has another symmetry. It is generated by
\eqn\Jflat{ {\cal H} = - \ha P_m P^m ,}
which determines that the only world-sheet field that varies is $X^m$ and transforms as $\d X^m = - \e P^m$, where $\e$ is a conformal weight $-1$ parameter,  and the action is invariant. It is trivial that ${\cal H}$ is BRST invariant because $P_m$ is BRST invariant. Note that \Jflat\ acting on physical states provides the massless condition explicitly, although this is not necessary in flat space because the cohomology of \brst\ contains massless states only \BerkovitsXBA. 

In summary, we have a nilpotent BRST charge, and BRST invariant symmetries generated by ${\cal K}$ and ${\cal H}$. In the next section we generalize this to a curved supergravity background. 

\newsec{Curved Background}

The action in curved background is the covariantization of \ssusy, that is
\eqn\scurved{S = \int d^2z ~ P_a \Pib^a + d_\a \Pib^\a + \dhh_\ah \Pib^\ah + \o_\a \Nb \l^\a + \oh_\ah \Nb \lh^\ah ,} 
where $\Pib^A = \pb Z^M E_M{}^A$, with $E_M{}^A$ being the vielbein superfield, and $Z^M$ are the coordinates of the curved ten-dimensional superspace. The covariant derivatives are defined with the background Lorentz connections, that is,
\eqn\cov{Ê\Nb \l^\a = \pb \l^\a + \l^\b \Ob_\b{}^\a,\quad \Nb \lh^\ah = \pb \lh^\ah + \lh^\bh \Ob_\bh{}^\ah ,} 
where $\Ob_\b{}^\a = \pb Z^M \O_{M\b}{}^\a$ and $\Ob_\bh{}^\ah = \pb Z^M \Oh_{M\bh}{}^\ah$. 

The equations of motion from the action \scurved\ are 
\eqn\eomc{Ê\pb Z^M = 0,\quad \pb \l^\a = 0,\quad \pb \lh^\ah = 0,\quad \pb \o_\a = 0,\quad \pb \oh_\ah = 0 .} 
$$\pb d_\a = 0,\quad \pb \dhh_\ah = 0 .$$
That is, all the world-sheet variables in \scurved\ are holomorphic. 

The BRST charge is given by \brst. Because of \eomc\ the BRST charge is conserved and we do not obtain any constraint from this as opposed to the usual string where the conservation of the BRST charge determines a set of constraints on the background fields \BerkovitsUE. 

We now determine the constraints determined by the nilpotency of the BRST charge. To obtain them, we first compute the BRST transformations of the world-sheet fields of the action \scurved.  The idea is to express the world-sheet fields $d_\a$ and $\dhh_\ah$ in terms of conjugate variables, $(\l, \o)$ and $(\lh, \oh)$ are already conjugate pairs. We define the momentum of $Z^M$ and then use canonical commutation relations as in \BerkovitsUE. The momentum for $Z^M$ is
\eqn\defP{P_M = {{\d S}\over{\d (\p_\tau Z^M )}} = E_M{}^a P_a - E_M{}^\a d_\a - E_M{}^\ah \dhh_\ah + \O_{M\a}{}^\b \l^\a \o_\b + \Oh_{M\ah}{}^\bh \lh^\ah \oh_\bh ,} 
and the commutation relation is\foot{We relate $(z,\bar{z}) \to (\tau,\s)$ such that $\p = \p_\tau - \p_\s, \pb = \p_\tau + \p_\s$. }
\eqn\comPZ{[ P_M(\s) , Z^N(\s') ] = - \d_M^N \d(\s-\s') ,} 
 at equal world-sheet times.  Simillalry, the other canonical commutation relations in \scurved\ are
 \eqn\compure{[ \l^\a (\s) , \o_\b(\s') ] = \d_\b^\a \d(\s-\s'),\quad [Ê\lh^\ah(\s) , \oh_\bh(\s') ] = \d_\bh^\ah \d(\s-\s') ,} at equal world-sheet times.  
 
The BRST transformation of any field $\Psi$ is defined as
\eqn\defbrst{ÊQ \Psi(\s) = \oint d\s' [ \l^\a d_\a(\s') + \lh^\ah \dhh_\ah(\s') , \Psi(\s) ] ,}
and after expressing the integrand of $Q$ in terms of conjugate variables and using the commutation relations \comPZ\ and \compure, the BRST transformations can be obtained as it was done in the heterotic pure spinor string \ChandiaIX\ and in the type II pure spinor superstring \ChandiaSTA,  both in a curved background.

The BRST transformation of the superspace variable $Z^M$ is
\eqn\QZ{ÊQ Z^M = \l^\a E_\a{}^M + \lh^\ah E_\ah{}^M ,}
and the pure spinor variables transform as
\eqn\Qlo{ÊQ\l^\a = -\l^\b ( \l^\g \O_{\g\b}{}^\a + \lh^\gh \O_{\gh\b}{}^\a ),\quad Q\lh^\ah  = -\l^\bh ( \l^\g \Oh_{\g\bh}{}^\ah + \lh^\gh \Oh_{\gh\bh}{}^\ah ) ,}
$$ Q\o_\a = d_\a + ( \l^\g \O_{\g\a}{}^\b + \lh^\gh \O_{\gh\a}{}^\b ) \o_\b,\quad Q\oh_\ah = \dhh_\ah + ( \l^\g \Oh_{\g\ah}{}^\bh + \lh^\gh \Oh_{\gh\ah}{}^\bh ) \oh_\bh ,$$ 
note that the terms depending on $\O$ and $\Oh$ are Lorentz rotations with field-dependent parameters.

Using
\eqn\ddhP{Êd_\a = -E_\a{}^M P_M + \O_{\a\b}{}^\g \l^\b \o_\g + \Oh_{\a\bh}{}^\gh \lh^\bh \oh_\gh ,} 
$$ \dhh_\ah = - E_\ah{}^M P_M + \O_{\ah\b}{}^\g \l^\b \o_\g + \Oh_{\ah\bh}{}^\gh \lh^\bh \oh_\gh ,$$
$$ P_a = E_a{}^M P_M - \O_{a\a}{}^\b \l^\a \o_\b - \Oh_{a\ah}{}^\bh \lh^\ah \oh_\bh ,$$
we obtain
\eqn\Qd{ Q d_\a = \l^\b \left( T_{\b\a}{}^a P_a - T_{\b\a}{}^\g d_\g - T_{\b\a}{}^\gh \dhh_\gh \right) + \lh^\bh \left( T_{\bh\a}{}^a P_a - T_{\bh\a}{}^\g d_\g - T_{\bh\a}{}^\gh \dhh_\gh \right) }
$$+ \l^\b \left( \l^\g \o_\d R_{\a\b\g}{}^\d + \lh^\gh \oh_\dh R_{\a\b\gh}{}^\dh \right) + \lh^\bh \left( \l^\g \o_\d R_{\a\bh\g}{}^\d + \lh^\gh \oh_\dh R_{\a\bh\gh}{}^\dh \right) $$
$$  + \left( \l^\g \O_{\g\a}{}^\b + \lh^\gh \O_{\gh\a}{}^\b \right) d_\b ,$$  
\eqn\Qdh{ÊQ \dhh_\ah = \l^\b \left( T_{\b\ah}{}^a P_a - T_{\b\ah}{}^\g d_\g - T_{\b\ah}{}^\gh \dhh_\gh \right) + \lh^\bh \left( T_{\bh\ah}{}^a P_a - T_{\bh\ah}{}^\g d_\g - T_{\bh\ah}{}^\gh \dhh_\gh \right) }
$$+ \l^\b \left( \l^\g \o_\d R_{\ah\b\g}{}^\d + \lh^\gh \oh_\dh R_{\ah\b\gh}{}^\dh \right) + \lh^\bh \left( \l^\g \o_\d R_{\ah\bh\g}{}^\d + \lh^\gh \oh_\dh R_{\ah\bh\gh}{}^\dh \right) $$
$$  + \left( \l^\g \Oh_{\g\ah}{}^\bh + \lh^\gh \Oh_{\gh\ah}{}^\bh \right) \dhh_\bh ,$$   
\eqn\QP{ Q P_a = \l^\a \left( - T_{\a a}{}^b P_b + T_{\a a}{}^\b d_\b + T_{\a a}{}^\bh \dhh_\bh \right) + \lh^\ah \left( - T_{\ah a}{}^b P_b + T_{\ah a}{}^\b d_\b + T_{\ah a}{}^\bh \dhh_\bh \right) } 
$$ + \l^\a \left( \l^\b \o_\g R_{a\a\b}{}^\g + \lh^\bh \oh_\gh R_{a\a\bh}{}^\gh \right) + \lh^\ah \left( \l^\b \o_\g R_{a\ah\b}{}^\g + \lh^\bh \oh_\gh R_{a\ah\bh}{}^\gh \right) $$
$$ +\left( \l^\a \O_{\a a}{}^b + \lh^\ah \O_{\ah a}{}^b \right) P_b .$$
As in \Qlo, there are Lorentz rotation pieces in these transformations. The torsion and curvatures are such that $T_{AB}{}^\a$ and $R_{AB\a}{}^\b$ are defined with the connection $\O$, and $T_{AB}{}^\ah$ and $R_{AB\ah}{}^\bh$ are defined with the connection $\Oh$. We follow the conventions for superspace differential forms used in \BedoyaIC. 

For future reference, we find the BRST transformation of $\Pi^A$ and the BRST transformation of the connections in \cov. Because $Z^M$ has the same BRST transformation as the minimal pure spinor string, the $\Pi^A$ transforms similarly \ChandiaSTA,
\eqn\Qpi{ Q\Pi^a = -\l^\a \Pi^A T_{A\a}{}^a - \lh^\ah \Pi^A T_{A\ah}{}^a - \Pi^b ( \l^\a \O_{\a b}{}^a + \l^\ah \O_{\ah b}{}^a ) ,}Ê
$$ Q \Pi^\a = \N \l^\a - \l^\b \Pi^A T_{A\b}{}^\a - \lh^\bh \Pi^A T_{A\bh}{}^\a + \Pi^\b ( \l^\g \O_{\g\b}{}^\a + \lh^\gh \O_{\gh\b}{}^\a ) ,$$
$$ Q \Pi^\ah = \N \lh^\ah - \l^\b \Pi^A T_{A\b}{}^\ah - \lh^\bh \Pi^A T_{A\bh}{}^\ah + \Pi^\bh  (\l^\g \Oh_{\g\bh}{}^\ah + \lh^\gh \Oh_{\gh\bh}{}^\ah ) .$$
As in \Qlo and \Qd-\QP, the last terms here correspond to a Lorentz rotation.

The BRST transformations of the connections can be derived from $Q Z^M$, the result is
\eqn\QOM{ÊQ \O_\a{}^\b = - \l^\g \Pi^A R_{A\g\a}{}^\b - \lh^\gh \Pi^A R_{A\gh\a}{}^\b + \N (\l^\g \O_{\g\a}{}^\b + \lh^\gh \O_{\gh\a}{}^\b )   ,}
$$ Q \O_\ah{}^\bh = - \l^\g \Pi^A R_{A\g\ah}{}^\bh - \lh^\gh \Pi^A R_{A\gh\ah}{}^\bh + \N (\l^\g \Oh_{\g\ah}{}^\bh + \lh^\gh \Oh_{\gh\ah}{}^\bh ) , $$
where the last terms are, again, a Lorentz rotation.

We can compute $Q^2$ using the action of $Q$ on itself. We obtain
\eqn\QQ{ÊQ^2 = \oint \l^\a \l^\b ( ~T_{\a\b}{}^a P_a - T_{\a\b}{}^\g d_\g - T_{\a\b}{}^\gh \dhh_\gh + \l^\g \o_\d R_{\a\b\g}{}^\d + \lh^\gh \oh_\dh R_{\a\b\gh}{}^\dh ~ ) }  
 $$+ \oint \lh^\ah \lh^\bh  ( ~T_{\ah\bh}{}^a P_a - T_{\ah\bh}{}^\g d_\g - T_{\ah\bh}{}^\gh \dhh_\gh + \l^\g \o_\d R_{\ah\bh\g}{}^\d + \lh^\gh \oh_\dh R_{\ah\bh\gh}{}^\dh ~ ) $$
$$+ \oint \l^\a \lh^\bh  ( ~T_{\a\bh}{}^a P_a - T_{\a\bh}{}^\g d_\g - T_{\a\bh}{}^\gh \dhh_\gh + \l^\g \o_\d R_{\a\bh\g}{}^\d + \lh^\gh \oh_\dh R_{\a\bh\gh}{}^\dh ~ ) .$$ 
 From $Q^2 = 0$ we obtain the nilpotency constraints
\eqn\QQCone{ \l^\a \l^\b T_{\a\b}{}^A = \l^\a \l^\b R_{\a\b\gh}{}^\dh = \l^\a \l^\b \l^\g R_{\a\b\g}{}^\d = 0 ,}
\eqn\QQCtwo{  \lh^\ah \lh^\bh T_{\ah\bh}{}^A = \lh^\ah \lh^\bh R_{\ah\bh\g}{}^\d = \lh^\ah \lh^\bh \lh^\gh R_{\ah\bh\gh}{}^\dh = 0 ,}
\eqn\QQCthree{ÊT_{\a\bh}{}^A = \l^\a \l^\b R_{\gh\a\b}{}^\d = \lh^\ah \lh^\bh R_{\g\ah\bh}{}^\dh = 0 ,} 
which are the constraints from nilpotency of the type II superstring BRST operator in \BerkovitsUE\ involving torsion and curvature components. Note that the action the action \scurved\ is identically invariant under BRST transformations and it does not provide new constraints on background fields.

We will prove now that the remaining nilpotency constraints of \BerkovitsUE\ come from BRST invariance of the curved background generalization of \qflat, that is
\eqn\qcurved{Ê{\cal K} = \oint \l^\a \Pi^A B_{A\a} + \lh^\ah \Pi^A B_{A\ah}Ê= \oint \l^\a \p Z^M B_{M\a} + \lh^\ah \p Z^M B_{M\ah} .}
We will first prove that ${\cal K}$ is invariant, up to a BRST term, under 
\eqn\dB{ \d B_{NM} = \p_{[N} \L_{M]} ,}
where the parameter $\L_M$ allows to define the $(1,0)$ world-sheet form $\L = \p Z^M \L_M$. Note that this is required to have a theory invariant under this gauge symmetry. Varying  ${\cal K}$ under \dB\ we obtain
\eqn\dkone{Ê\d {\cal K}Ê= \oint -(-1)^M \l^\a \p Z^M \p_\a \L_M + \l^\a \p \L_\a - \l^\a \p Z^N (\p_N E_\a{}^M) \L_M }
$$+\oint -(-1)^M \lh^\ah \p Z^M \p_\ah \L_M + \lh^\ah \p \L_\ah - \lh^\ah \p Z^N (\p_N E_\ah{}^M) \L_M .$$
Note that the first terms in both lines combine to produce a term with $\l^\a \p_\a \L_M + \lh^\ah \p_\ah \L_M = Q \L_M$, then
\eqn\dktwo{ÊÊ\d {\cal K}Ê= \oint -(-1)^M \p Z^M Q \L_M + \l^\a \p \L_\a + \lh^\ah \p \L_\ah - \l^\a \p Z^N (\p_N E_\a{}^M) \L_M - \lh^\ah \p Z^N (\p_N E_\ah{}^M) \L_M }
$$= \oint - Q ( \p Z^M  \L_M ) + \p( Q Z^M ) \L_M  + \l^\a \p \L_\a + \lh^\ah \p \L_\ah - \l^\a \p Z^N (\p_N E_\a{}^M) \L_M - \lh^\ah \p Z^N (\p_N E_\ah{}^M) \L_M $$
$$ = \oint - Q ( \p Z^M  \L_M )  + \p( \l^\a \L_\a + \lh^\ah \L_\ah ) = - Q \oint \L .$$ 
Then, we have proved that the variation of \qcurved\ is invariant under the gauge symmetry of \dB\ up to a BRST-exact term. 
 
We now show that the BRST invariance of \qcurved\ implies the nilpotency constraints of \BerkovitsUE\ involving $H=dB$. Using \Qlo\ and \Qpi\ we obtain 
\eqn\QKC{ÊQ{\cal K} = - \oint \ha \l^\a \l^\b \Pi^A ( H_{A\b\a} - T_{\b\a}{}^B B_{BA} ) +}
$$ \ha \lh^\ah \lh^\bh \Pi^A ( H_{A\bh\ah}Ê- T_{\bh\ah}{}^B B_{BA} ) + \l^\a \lh^\bh \Pi^A ( H_{A\bh\a} - T_{\bh\a}{}^B B_{BA} ) ,$$  
and after using \QQCone-\QQCthree\ we obtain the nilpotency constraints of \BerkovitsUE\ for $H$,
\eqn\QQH{Ê\l^\a \l^\b H_{\a\b A} = \lh^\ah \lh^\bh H_{\ah\bh A} = H_{\a\bh A} = 0 .}

We can obtain the transformations of the fields in \scurved\ under $\KK$. The fields $Z^M, \l^\a, \lh^\ah$ are invariant. It is easy to obtain the $\KK$-transformations of $P_M, \o_\a, \oh_\ah$, they are
\eqn\Kooh{Ê\KK \o_\a = \p Z^M B_{M\a} = \Pi^A B_{A\a},\quad \KK \oh_\ah = \p Z^M B_{M\ah} = \Pi^A B_{M\ah} ,} 
\eqn\KPM{ \KK P_M = \p\l^\a B_{\a M}Ê+ \p\lh^\ah B_{\ah M} - (-1)^M \l^\a \p Z^N \p_{[N} B_{M]\a}  - (-1)^M \lh^\ah \p Z^N \p_{[N} B_{M]\ah} .} 
Using these transformations and \ddhP\ we obtain the $\KK$-transformations of $d_\a, \dhh_\ah, P_a$, they are
\eqn\Kda{ \KK d_\a = - \l^\b \Pi^a H_{a\b\a} - \l^\b \Pi^\g H_{\g\b\a} - Q b_\a - \l^\b T_{\b\a}{}^A b_A + ( \l^\g \O_{\g\a}{}^\b + \lh^\gh \O_{\gh\a}{}^\b ) b_\b ,}
\eqn\Kdhah{ \KK \dhh_\ah = - \lh^\bh \Pi^a H_{a\bh\ah} - \lh^\bh \Pi^\gh H_{\gh\bh\ah} - Q b_\ah - \lh^\bh T_{\bh\ah}{}^A b_A + ( \l^\g \Oh_{\g\ah}{}^\bh + \lh^\gh \Oh_{\gh\ah}{}^\bh ) b_\bh ,}
\eqn\KPa{ \KK P_a =  \l^\b \Pi^b H_{\b ab} + \l^\b \Pi^\g H_{\g\b a} + \l^\bh \Pi^b H_{\bh ab}Ê+ \lh^\bh \Pi^\gh H_{\gh\bh a}  + Q b_a $$ $$+ ( \l^\b T_{\b a}{}^A + \lh^\bh T_{\bh a}{}^A ) b_A - ( \l^\g \O_{\g a}{}^b + \lh^\gh \O_{\gh a}{}^ b ) b_b ,}
where $b_A = \Pi^B B_{BA}$.  It turns out that the action \scurved\ is invariant under $\KK$. The transformation of the action is
\eqn\KSC{Ê\KK S = \int d^2z ~ \l^\a \Pi^A \Pib^B H_{BA\a} + \lh^\ah \Pi^A \Pib^B H_{BA\ah} - Q ( \Pi^A \Pib^B B_{BA} ) ,}
which vanishes because the BRST transformation of the second term cancels the first term. Then, the invariance of the action under $\KK$-transformations does not provide new constraints on background fields.

We now generalize \Jflat\ to get the other supergravity constraints of \BerkovitsUE. We propose that 
\eqn\HC{ {\cal H} = -\ha P_a P_b \eta^{ab} + d_\a \dhh_\bh P^{\a\bh} + d_\a \lh^\bh \oh_\gh \Ch_\bh{}^{\gh\a} + \dhh_\ah \l^\b \o_\g C_\b{}^{\g\ah} + \l^\a \o_\b \lh^\gh \oh_\dh S_{\a\gh}{}^{\b\dh} ,}
is the curved background version of \Jflat. Here , $P^{\a\bh}$ is a superfield whose lowest components are the Ramond-Ramond field-strength, $C_\b{}^{\g\ah}$ and $\Ch_\bh{}^{\gh\a}$ are superfields whose lowest components are related to gravitini and dilatini field-strengths  and $S_{\a\gh}{}^{\b\dh}$ is a superfield whose lowest component is the space-time curvature \BerkovitsUE. Note that all the terms in \HC\ should respect the pure spinor gauge symmetries $\d \o_\a = (\g^m\l)_\a \L_m$ and $\d \oh_\ah = (\g^m \lh)_\ah \L_m'$. 

Defining $P^a = \eta^{ab} P_b$ and using the BRST transformations determined above, we obtain
\eqn\qjota{
Q {\cal H} = \ha \l^\a P^a P^b  T_{\a(ab)} + \ha \lh^\ah P^a P^b T_{\ah(ab)} + \l^\a P^a d_\b  T_{a\a}{}^\b + \lh^\ah P^a  \dhh_\bh T_{a\ah}{}^\bh } 
$$+ \lh^\ah P^a d_\b ( T_{a\ah}{}^\b - T_{\ah\gh a} P^{\b\gh} )
 + \l^\a P^a \dhh_\bh ( T_{a\a}{}^\bh + T_{\a\g a} P^{\g\bh} ) $$
 $$ - \l^\a P^a \l^\b \o_\g R_{a\a\b}{}^\g + \l^\ah P^a \l^\b \o_\g  ( -R_{a\ah\b}{}^\g + T_{\ah\dh a} C_\b{}^{\g\ah} ) $$
$$ + \l^\a P^a \lh^\bh \oh_\gh ( -R_{a\a\bh}{}^\gh + T_{\a\d a} \Ch_\bh{}^{\gh\d} ) - \l^\ah P^a \lh^\bh \oh_\gh R_{a\ah\bh}{}^\gh + \ha \lh^\ah d_\b d_\g T_{\ah\dh}{}^{[\b} P^{\g]\dh}  - \ha \l^\a \dhh_\bh \dhh_\gh T_{\a\d}{}^{[\bh} P^{\d \bh]} $$ 
$$+ \l^\a d_\b \dhh_\gh ( \N_\a P^{\b\gh} - T_{\a\d}{}^\b P^{\d\gh} + C_\a{}^{\b\gh} ) + \lh^\ah d_\b \dhh_\gh ( \N_\ah P^{\b\gh} - T_{\ah\dh}{}^\gh P^{\b\dh} - \Ch_\ah{}^{\gh\b} ) $$
$$+ \lh^\ah d_\b \l^\g \o_\d ( - R_{\rh\ah\g}{}^\d P^{\b\rh} + T_{\ah\rh}{}^\b C_\g{}^{\d\rh} ) + \l^\a d_\b \lh^\gh \oh_\dh ( - R_{\rh\a\gh}{}^\dh P^{\b\rh} + T_{\a\r}{}^\b \Ch_\gh{}^{\dh\r} - \N_\a \Ch_\gh{}^{\dh\b} + S_{\a\gh}{}^{\b\dh} ) $$
$$- \lh^\ah d_\b \lh^\gh \oh_\dh ( \N_\ah \Ch_\gh{}^{\dh\b} + R_{\rh\ah\gh}{}^\dh P^{\b\rh} ) - \l^\a \dhh_\bh \l^\g \o_\d ( \N_\a C_\g{}^{\d\bh} - R_{\r\a\g}{}^\d P^{\r\bh} ) $$ 
$$+ \lh^\ah \dhh_\bh \l^\g \o_\d ( R_{\r\ah\g}{}^\d P^{\r\bh} + T_{\ah\rh}{}^\bh C_\g{}^{\d\rh} - \N_\ah C_\g{}^{\d\bh} + S_{\g\ah}{}^{\d\bh} ) + \l^\a \dhh_\bh \lh^\gh \oh_\dh ( R_{\r\a\bh}{}^\gh P^{\r\bh} + T_{\a\r}{}^\bh \Ch_\gh{}^{\dh\r} )$$ 
$$+ \l^\a \l^\b \o_\g \lh^\dh \oh_\rh ( R_{\s\a\b}{}^\g \Ch_\dh{}^{\rh\s} + R_{\sh\a\dh}{}^\rh C_\b{}^{\g\sh} + \N_\a S_{\b\dh}{}^{\g\rh} ) $$ $$+ \lh^\ah \l^\b \o_\g \lh^\dh \oh_\rh ( R_{\s\ah\b}{}^\g \Ch_\dh{}^{\rh\s} + R_{\sh\ah\dh}{}^\rh C_\b{}^{\g\sh} + \N_\ah S_{\b\dh}{}^{\g\rh} ) .$$
The BRST invariance of ${\cal H}$ implies the constraints
\eqn\qjone{ T_{\a(ab)} = T_{a\a}{}^\b = T_{a\a}{}^\bh + T_{\a\g a} P^{\g\bh} = \l^\a \l^\b R_{a\a\b}{}^\g = 0 ,}
\eqn\qjtwo{ T_{\ah(ab)} = T_{a\ah}{}^\bh = T_{a\ah}{}^\b - T_{\ah\gh a} P^{\b\gh} = \lh^\ah \lh^\bh R_{a\ah\bh}{}^\gh = 0 ,}
\eqn\qjthree{ R_{a\ah\b}{}^\g - T_{\ah\dh a} C_\b{}^{\g\ah} = R_{a\a\bh}{}^\gh - T_{\a\d a} \Ch_\bh{}^{\gh\d} = 0 ,}
\eqn\qjfour{\N_\a P^{\b\gh} - T_{\a\d}{}^\b P^{\d\gh} + C_\a{}^{\b\gh} = \N_\ah P^{\b\gh} - T_{\ah\dh}{}^\gh P^{\b\dh} - \Ch_\ah{}^{\gh\b} = 0 ,}
\eqn\qjfive{ \N_\a \Ch_\gh{}^{\dh\b} - T_{\a\r}{}^\b \Ch_\gh{}^{\dh\r} + R_{\rh\a\gh}{}^\dh P^{\b\rh} - S_{\a\gh}{}^{\b\dh} = 0,}
\eqn\qjsix{ \N_\ah C_\g{}^{\d\bh} - T_{\ah\rh}{}^\bh C_\g{}^{\d\rh} - R_{\r\ah\g}{}^\d P^{\r\bh} - S_{\g\ah}{}^{\d\bh} = 0 ,}
\eqn\qjseven{ \l^\a \l^\b ( \N_\a C_\b{}^{\g\dh} - R_{\r\a\b}{}^\g P^{\r\dh} ) = \lh^\ah \lh^\bh ( \N_\ah \Ch_\bh{}^{\d\gh} + R_{\rh\ah\bh}{}^\gh P^{\d\rh} ) = 0 ,}
\eqn\qjeight{ \l^\a \l^\b  ( R_{\s\a\b}{}^\g \Ch_\dh{}^{\rh\s} + R_{\sh\a\dh}{}^\rh C_\b{}^{\g\sh} + \N_\a S_{\b\dh}{}^{\g\rh} ) = 0 ,}
\eqn\qjnine{ \lh^\ah \lh^\gh ( R_{\s\ah\b}{}^\g \Ch_\dh{}^{\rh\s} + R_{\sh\ah\dh}{}^\rh C_\b{}^{\g\sh} + \N_\ah S_{\b\dh}{}^{\g\rh} ) = 0 .}
Note that this gives the holomorphic constraints of Berkovits and Howe that do not involve $H=dB$ \BerkovitsUE. We also obtain the constraints

\eqn\newone{ T_{\ah\dh}{}^{[\b} P^{\g]\dh} = 0 ,}
\eqn\newtwo{ T_{\a\d}{}^{[\bh} P^{\d\gh]} = 0 ,}
\eqn\newthree{ R_{\rh\ah\g}{}^\d P^{\b\rh} - T_{\ah\rh}{}^\b C_\g{}^{\d\rh} = 0 ,}
\eqn\newfour{ R_{\r\a\gh}{}^\dh P^{\r\bh} + T_{\a\r}{}^\bh \Ch_\gh{}^{\dh\r} = 0 ,}
which are implied by the constraints of \BerkovitsUE. 

What remains to obtain are the constraints coming from the conservation of the BRST charge that involve $H=dB$ in \BerkovitsUE. They are,
\eqn\bhforH{ H_{\a ab} = H_{\ah ab} = T_{\a\b a} - H_{\a\b a} = T_{\ah\bh a} + H_{\ah\bh a} = 0 ,}
\eqn\bhforHone{ T_{\g\a}{}^\bh + \ha P^{\d\bh} H_{\d\g\a} = T_{\gh\ah}{}^\b + \ha P^{\b\dh} H_{\gh\dh\ah} = 0 ,}
\eqn\bhforHtwo{ R_{\d\a\bh}{}^\gh + \ha \Ch_\bh{}^{\gh\r} H_{\r\d\a} = R_{\dh\ah\b}{}^\g - \ha C_\b{}^{\g\rh} H_{\rh\dh\ah} = 0.} 
Note that \bhforHtwo\ are implied by the other constraints through Bianchi identities. Using the constraints \bhforHone\ and \bhforHtwo\ and the conditions from $Q\KK=0$ the equations \newone-\newfour\ are satisfied. This, however, does not imply we derived the 
constraints \bhforHone\ and \bhforHtwo. We will comment on this at the end of the paper. 

\newsec{ $AdS_5 \times S^5$ Background}

The $AdS$ geometry is described by the supercoset $PSU(2,2|4)/SO(1,4)\times SO(5)$. The background fields 
are conveniently described using the Maurer-Cartan currents
\eqn\Js{ J_M = g^{-1}\partial_M g=(g^{-1}\partial_M g)^{[\underline{ab}]}T_{[\underline{ab}]}+(g^{-1}\partial_M g)^A T_{A} ,}
where $g$ is an element of the coset and $\{T_{[\underline{ab}]},T_A\}$ are the generators of the $\frak{psu}(2,2|4)$ algebra, $T_{[\underline{ab}]}$ generates the $SO(1,4)\times SO(5)$ algebra. We will also define 
$$J= \partial Z^M g^{-1}\partial_M g,\quad \Jb = \pb Z^M g^{-1}\partial_M g .$$

The $\frak{psu}(2,2|4)$ has a ${\Bbb Z}_4$ symmetry and the generators can be conveniently grouped 
in four different subspaces labeled by the different ${\Bbb Z}_4$ charges
$$ J = J_0 + J_1 +J_2 +J_3 = \Omega^{[\underline{ab}]}T_{[\underline{ab}]} + J^\alpha T_\alpha + 
J^{\underline{a}}T_{\underline{a}} + J^\ah T_\ah .$$ 
We can now identity the vielbein and spin connection 
$$  \Omega_M{}^{[\underline{ab}]} = (g^{-1}\partial_M g)^{[\underline{ab}]},\quad E_M{}^A=(g^{-1}\partial_M g)^A.$$

In the case of $AdS_5\times S^5$ the previous general curved space action is 
\eqn\adsaction{ S= \int\! d^2z \, P_{\underline{a}} \Jb^{\underline{a}} + d_\alpha \Jb^{\alpha} + \dhh_\ah \Jb^\ah + \omega_\alpha \Nb \lambda^\alpha +\oh_\ah\Nb\lh^\ah .}
Introducing the notation 
\eqn\shortnt{P_2=P_{\underline{a}}T_{\underline{b}}\eta^{\underline{ab}},\quad P_3 = - d_\alpha T_\bh\eta^{\alpha\bh},\quad P_1 = \dhh_\ah T_\beta \eta^{\beta\ah} ,}
$$ \lambda = \lambda^\alpha T_\alpha, \quad \omega = - \omega_\alpha T_\bh\eta^{\alpha\bh},\quad \lh = \lh^\ah T_\ah,\quad \oh = \oh_\ah T_\beta \eta^{\beta\ah},$$
where $\eta^{AB}$ is the inverse of the Killing form $\eta_{AB}={\rm Str}(T_AT_B)$, we can write the action as 
\eqn\simpleact{S= \int\! d^2z\, {\rm Str}(P_2 \Jb_2 + P_3 \Jb_1 + P_1 \Jb_3 + \omega\Nb\lambda + \oh\Nb\lh),}
and the BRST charge as 
\eqn\BRSTAdS{Q = \oint {\rm Str}(\lambda P_3 + \lh P_1).}

The equations of motion that follow from this action are 
\eqn\EOMAdS{\Jb_1=\Jb_2=\Jb_3= \Nb P_1=\Nb P_2=\Nb P_3=0,}
\eqn\EOMAdSS{\Nb\lambda=\Nb\omega=\Nb\lh=\Nb\oh=0.}

Note that \EOMAdS\ does not imply $g$ is holomorphic. However, using the Maurer-Cartan identities \EOMAdS\ implies 
\eqn\EOMAdSSS{\Nb J_2 = \Nb J_1=\Nb J_3 =0 .}

Using the background field values of $P^{\alpha\bh}$ and $S_{\a\gh}{}^{\b\dh}$ the for  $AdS$ case, the constraint $\cal H$ is given by 
$${\cal H} = -{1\over 2}{\rm Str}( P_2P_2 +2 P_3P_1 +2 N\Nhh ),$$
where $N=\{\lambda,\omega\}$ and $\Nhh =\{\lh,\oh\}$. This constraint has precisely the expected form of a curved space laplacian. When acting on physical states, $\cal H$ is the quadratic Casimir of the representation of the vertex operator.  

For the $\cal K$ charge, we use the background field value of $B_{\alpha\bh}$ and the result is 
\eqn\ChargeK{ {\cal K} = \oint {\rm Str}(\lambda J_3 - \hat\lambda J_1).} 
Using the commutation relations obtained in the previous section, one can show that 
$Q{\cal K}=0$. It is interesting to note that ${\cal K}$ resembles the BRST charge of the usual pure spinor string in $AdS$, where $P_1$ and $P_3$ can be integrated out due to 
the constant Ramond-Ramond field-strength. In the ambitwistor case, this field-strength 
is not in the action but in $\cal H$. 

The global $\frak{psu}(2,2|4)$ symmetry of the usual string in $AdS_5\times S^5$ is promoted to an holomorphic transformation of the coset element
\eqn\HoloG{\delta g = \Lambda(z)g,\quad \delta P_i=\delta\lambda=\delta\omega=\delta\lh=\delta\oh=0.}
The conserved current that generates this transformation can be calculated from the action \simpleact\ using the standard Noether procedure and the result is 
\eqn\jholo{ j = g( P_2 + P_1 +P_3 +N+\Nhh)g^{-1}.}
This current is BRST invariant, as opposed to the usual case where the BRST transformation 
of the symmetry currents is an exact form \MikhailovRX. As discussed in \BerkovitsXBA\ the integrated vertex operator in flat space is BRST invariant, contrary to the expected $QV = d\Lambda$. Mikhailov has shown in \MikhailovRX\ that a class of massless vertex operators for the $AdS$ pure spinor string can be constructed using the symmetry currents. Given this relation, it is not surprising that the currents generating 
\HoloG\ are BRST invariant. 

Here we conjecture that the components of the current $j= j^{\underline A} T_{\underline A}$, where $T_{\underline A}$ are all $\frak{psu}(2,2|4)$ generators, satisfy a Kac-Moody algebra 
\eqn\KMope{ j^{\underline A}(z)j^{\underline B}(y)\to {\kappa^{\underline{AB}}\over (z-y)^2} + {{ f^{\underline{AB}}_{\underline C} j^{\underline C}(y) }\over (z-y)},}
where $\kappa^{\underline{AB}}$ still has to be determined and $f^{\underline{AB}}_{\underline C}$ are the $\frak{psu}(2,2|4)$ structure constants. The usual Sugawara construction  using $j$ gives precisely the $\cal H$ constraint  
\eqn\niceH{ {\cal H}= -{1\over 2}{\rm Str}(j^2).}
 
 We have used the pure spinor constraint to show that ${\rm Str}(N^2)={\rm Str}(\hat N^2)=0$. It is important to note that, as in the previous sections, $\cal H$ is not the stress-energy tensor, which is the conserved current of conformal transformations $\delta g = \epsilon(z) \partial g$ and is given by 
$$T= {\rm Str}( P_2 J_2 + P_3 J_1 +P_1 J_3 + \omega\nabla\lambda + \oh\nabla\lh).$$

\newsec{Concluding Remarks}

In this paper we have studied the pure spinor formulation in a type II supergravity background for the ambitwistor string. This string describes only the supergravity part of the usual superstring spectrum. As in \AdamoWEA\ we expect that the conditions on $Q$, ${\cal H}$ and ${\cal K}$ do not receive any quantum correction. However, the quantum version of these constraints could receive corrections. It would be interesting to verify this explicitly. 

The curved space action \scurved\ is much simpler than the usual pure spinor string. All conjugate momenta appear at most linearly and the equations of motion still imply all fields are holomorphic. These two facts indicates perturbative calculations are easier 
and, as in \AdamoWEA, could be computed exactly. 

As we pointed out in the main text, we do not obtain all supergravity constraints. These missing contraints could be in the correct curved space generalization of the commutator of 
$\KK$ and $\HH$
\eqn\KHflat{ [\KK, \HH] =  2( \l\g^m\p\t - \lh\g^m\p\th ) P_m  + Q(\p\t\g^m\t P_m).}

Since $\KK$ and $\HH$ are symmetries of the flat space action, the right hand sice of \KHflat\ also is. Using the curved space versions of these generators we obtain that 

\eqn\KHC{ [\KK ,\HH] = - P^a  ( \l^\a \Pi^A H_{A\a a} + \lh^\ah \Pi^A H_{A\ah a} )}
$$ - \l^\d \Pi^A H_{A\d\a} ( \dhh_\bh P^{\a\bh} + \lh^\bh \oh_\gh \Ch_\bh{}^{\gh\a} )  
- \lh^\dh \Pi^A H_{A\dh\ah} ( -d_\b P^{\b\ah}Ê+ \l^\b \o_\g C_\b{}^{\g\ah} )$$
$$+Q ( -b_a P^a - b_\a ( \dhh_\bh P^{\a\bh} + \lh^\bh \oh_\gh \Ch_\bh{}^{\gh\a} )  - b_\ah ( -d_\b P^{\b\ah}Ê+ \l^\b \o_\g C_\b{}^{\g\ah} ) ),$$
where $b_A = \Pi^B B_{BA}$.
If one uses the desired constraints \bhforH-\bhforHtwo\ and the constraints derived from $Q\HH=0$, the expression above can be written, up to BRST exact terms,  as  
\eqn\KHCp{ [\KK,\HH] = ( -\l^\a \Pi^\b T_{\a\b}{}^a + \lh^\ah \Pi^\bh T_{\ah\bh}{}^a ) P_a} 
$$+ ( \lh^\bh \Pi^a T_{a\bh}{}^\a + 2 \lh^\bh \Pi^\gh T_{\gh\bh}{}^\a ) d_\a + ( \l^\b \Pi^a T_{a\b}{}^\ah - 2 \l^\b \Pi^\g T_{\g\b}{}^\ah ) \dhh_\ah$$
$$ +( \lh^\gh \Pi^a R_{a\gh\a}{}^\b + 2 \lh^\gh \Pi^\dh R_{\dh\gh\a}{}^\b ) \l^\a \o_\b - ( \l^\g \Pi^a R_{a\g\ah}{}^\bh + 2 \l^\g \Pi^\d R_{\d\g\ah}{}^\bh ) \lh^\ah \oh_\bh\, .$$ 

As of yet, we do not have an explanation for why the form \KHCp\ is preferred over \KHC. A possible to solve this issue is to consider the sum of the two nilpotent charges $Q+\KK$ as the new BRST charge \NathanPrivate. This can be interpreted as a redefinition of the conjugate momenta 
\eqn\RedefMoment{d_\alpha \to d_\alpha + b_\alpha,\quad \dhh_\ah \to \dhh_\ah + b_\ah.}
One can go further and also redefine $P_a \to P_a - b_a$. After defining a new $\HH'$ using these new momenta, the commutator $[Q+\KK,\HH']=0$ could give the missing constraints. The problem with this approach is that flat space limit of the modified BRST charge is not equal to the BRST charge defined originally by Berkovits in \BerkovitsXBA. So this modification would imply in a further change in the vertex operators found in that paper.  We plan to return to this problem together with questions about the quantum version of the constraints.  

The case of type IIB supergravity in an $AdS_5 \times S^5$ background is also simpler than the usual string. As it was shown in section 4, the global $\frak{psu}(2,2|4)$ symmetry is enhanced to an holomorphic symmetry. This symmetry could be used to completely solve the theory. If unintegrated and integrated vertex operators can be constructed explicitly, following \ChandiaKJA, correlation functions could be computed exactly. 

While this work was being completed we learned that Adamo, Berkovits and Casali  have been working on similar problems \ref\ABC{T. Adamo, N. Berkovits, E. Casali, private communication.}.

\bigskip

\noindent
{\bf Acknowledgements:} We would like to thank Nathan Berkovits, Eduardo Casali and William D. Linch III for useful comments. We also thank Nathan Berkovits and William D. Linch III for reading an early draft. This work is partially supported by FONDECYT grants 1120263 and 1151409 and by a CONICYT grant number DPI20140115.

\listrefs
 
\end